\documentclass[prd,aps,showpacs,preprintnumbers,amsmath, amssymb]{revtex4-2}

\usepackage{graphics,epsfig,hyperref}
\usepackage{graphicx}
\usepackage{dcolumn}
\usepackage{bm}
\usepackage{epstopdf}
\usepackage{color}
\usepackage{subfigure}
\usepackage{diagbox} 
\usepackage{adjustbox}
\usepackage{multirow}
\usepackage{amsmath,amssymb,amsfonts}

\begin{document}
	\baselineskip=0.8 cm
	\title{\bf Stationary scalar clouds around a rotating BTZ-like black hole in the Einstein-bumblebee gravity }
	
	\author{Fangli Quan$^{1}$, Fengjiao Li$^{1}$, Qiyuan Pan$^{1,2}$\footnote{panqiyuan@hunnu.edu.cn}, Mengjie Wang$^{1}$\footnote{mjwang@hunnu.edu.cn}, and Jiliang Jing$^{1,2}$\footnote{jljing@hunnu.edu.cn}}
	
	\affiliation{$^{1}$ Department of Physics, Key Laboratory of Low Dimensional Quantum Structures and Quantum Control of Ministry of Education, Institute of Interdisciplinary Studies, and Synergetic Innovation Center for Quantum Effects and Applications, Hunan Normal University,  Changsha, Hunan 410081, China}
	\affiliation{$^{2}$ Center for Gravitation and Cosmology, College of Physical Science and Technology, Yangzhou University, Yangzhou 225009, China}
	
	\begin{abstract}
		\baselineskip=0.6 cm
		\begin{center}
			{\bf Abstract}
		\end{center}

We have studied stationary clouds of massive scalar fields around a rotating BTZ-like black hole in the Einstein-bumblebee gravity, by imposing the Robin type boundary conditions at the AdS boundary. We establish, by scanning the parameter space, the existence of \textit{fundamental} stationary scalar clouds ($i.e.$, the overtone number $n=0$). In particular, we observe that the Lorentz symmetry breaking parameter $s$ and the quantum number $k$ play an opposite role in determining scalar clouds, which indicates the existence of \textit{degenerate} scalar clouds. To illustrate the fact that scalar clouds may only be supported for the $n=0$ case, we have analyzed the impact of various parameters on scalar quasinormal modes. It is shown that the Lorentz symmetry breaking parameter $s$ does not change the superradiance condition, and superradiant instabilities only appear for the fundamental modes. Our work shows that the Lorentz symmetry breaking provides richer physics in stationary scalar clouds around black holes.
		
	\end{abstract}
	
	\maketitle
	\newpage
	\vspace*{0.2cm}
	
	\section{Introduction}\label{Sec.I}
	
	In 1971, Ruffini and Wheeler proposed the no-hair conjecture \cite{RuffiniWheeler1971}, which is one of the most important developments in black hole physics, stating that stationary black holes should be characterized by only three parameters: mass, charge and angular momentum \cite{Carter1973,MisnerTW}. This pioneering work has led to many investigations concerning the new hair of black holes. Based on the superradiance mechanism, Hod first analyzed the dynamics of a test massive scalar field surrounding Kerr black holes \cite{HodPRD2012,HodEPJC2013} and further extended to the case of Kerr-Newman black holes \cite{HodPRD2014,HodPRD2016}, and found out that there exist the stationary and regular scalar field configurations surrounding the realistic rotating black holes which are coined as stationary scalar clouds. Herdeiro  \emph{et al.} presented a family of solutions of Einstein's gravity minimally coupled to a complex massive scalar field, which describes asymptotically flat, spinning black holes with scalar hair \cite{Herdeiro:2014goa}, and made a thorough investigation of the scalar clouds, which are bound states at the threshold of the superradiant instability, due to neutral (charged) scalar fields around Kerr(-Newman) black holes \cite{HerdeiroPRD2014,Herdeiro:2015gia,HerdeiroPRD2015}. Sampaio \emph{et al.} considered the Proca fields on the background of Reissner-Nordstr\"{o}m black holes and obtained the marginal Proca cloud configurations \cite{SampaioPRD2014}, and the authors of Ref. \cite{WangPRD2016} completed the analysis of bosonic fields on Kerr-anti-de Sitter (Kerr-AdS) black holes by studying Maxwell perturbations and constructed stationary Maxwell clouds with Robin boundary conditions. Extending the investigation to the excited states, Wang \emph{et al.} found that, in contrast to Kerr black holes with ground state scalar hair, the first excited Kerr black holes with scalar hair have two types of nodes, including radial $n_{r}=1$ and angular $n_{\theta}=1$ nodes \cite{Wang:2018xhw}. For a simple, geometrically elegant, black hole solution of three dimensional general relativity, Ferreira \emph{et al.} showed that the BTZ black hole can support stationary scalar clouds of a massive scalar field by using appropriate Robin boundary conditions \cite{Ferreira:2017cta,Dappiaggi:2017pbe}. Along this line, there have been accumulated interest to study various cloud configurations around black holes in the Einstein gravity \cite{HerdeiroRRPLB,OkawaCQG,HuangLiu,HodJHEP2017,GarciaSalgado,HodEPJC2020,LiuLPMPLA,GarciaSalgadoPRD2020,GarciaSalgadoPRD2021,HodPRD2023,GarciaSalgadoPRD2023,HodPRD2024,GuoWWY} and modified gravity \cite{LiEPJC,BenoneCHRPRD,RaduTYPRD,BernardPRD,HuangLZL,TokgozSakalli,GrandiLandea,Qiao:2020fta,HodPRD2021,CiszakMarino,HuangZhang,SiqueiraRichartz,ZhangEPJC2023,HodEPJC2023}.
	
	In this work, we focus on the stationary scalar clouds around a black hole in the Einstein-bumblebee gravity, which is one of simple effective theories with the Lorentz violation \cite{lvbh2}. In the Einstein-bumblebee model, the Lorentz violation arises from the dynamics of the bumblebee vector field $B_\mu$ \cite{MaiXLSPRD2023,ZhangWJSCPMA2023}. Casana \emph{et al.} found the first black hole solution in the Einstein-bumblebee gravity, $i.e.$, the exact Schwarzschild-like black hole solution, and presented the effects of the spontaneous Lorentz symmetry breaking on some classic tests, including the advance of perihelion, the bending of light, and Shapiro's time delay \cite{CasanaPRD2018}. More recently, Ding \emph{et al.} obtained a three-dimensional rotating BTZ-like black hole solution in the Einstein-bumblebee gravity theory, and found the effects of the Lorentz breaking constant on its thermodynamics and central charges \cite{Ding:2023niy}. Considering a scalar perturbation around such a rotating BTZ-like black hole in the Einstein-bumblebee gravity, Chen \emph{et al.} investigated the quasinormal modes of this rotating BTZ-like black hole and pointed out that the Lorentz symmetry breaking parameter imprints only in the imaginary parts of the quasinormal frequencies for the right-moving and left-moving modes \cite{ChenPLB2023}. Moreover, Ge \emph{et al.} constructed the mass ladder operators for the static BTZ-like black hole in the Einstein-bumblebee gravity, and probed the quasinormal frequencies of the mapped modes by the mass ladder operators for a scalar perturbation under the Dirichlet and Neumann boundary conditions \cite{GeCPC2025}. Since the three-dimensional gravity certainly offers potential insights into quantum gravity, it is of great interest to generalize the investigation to the stationary scalar clouds around a rotating BTZ-like black hole in the Einstein-bumblebee gravity. We will examine the influence of the spontaneous Lorentz symmetry breaking on the stationary scalar clouds, and check whether it is possible to distinguish between the BTZ black hole in the Einstein gravity and BTZ-like black hole in the Einstein-bumblebee gravity based on the scalar field configurations. 
	
	This work is organized as follows. In Section II, we briefly review the rotating BTZ-like black holes in the Einstein-bumblebee gravity, and present the massive Klein-Gordon equation and the appropriate boundary conditions for the non-extremal BTZ-like black holes. In Section III, we analyze stationary scalar configurations around the rotating BTZ-like black hole in the Einstein-bumblebee gravity. In Section IV, we investigate  quasinormal modes for a massive scalar field which is governed by the Klein-Gordon equation. We conclude in the last section with our main results. For completeness, in Appendix A, we study massive Klein-Gordon equation in the \textit{extremal} BTZ-like black hole.

	\section{Background geometry, massive scalar fields and boundary conditions}\label{Sec.II}

	\subsection{Rotating BTZ-like black holes}
	
	We start with the action of the Einstein-bumblebee gravity with a negative cosmological constant $\Lambda=-1/\ell^2$ in three dimensions \cite{string1,lvbh2,lvbh2s1,lvbhh1,lvbhh2,lvbhh12,lvbhh3,lvbhh4,lvbhh5,lvbhh6,lvbhh7,lvbhh8,lvbh10}
	\begin{eqnarray}\label{action}
		S=\int d^3x \sqrt{-g}\left[\frac{R-2\Lambda}{2\kappa} + \frac{\xi}{2\kappa} B^{\mu} B^{\nu}R_{\mu \nu}-\frac{1}{4} B^{\mu\nu} B_{\mu\nu} - V\left( B_{\mu} B^{\mu} \pm b^2 \right)\right],
	\end{eqnarray}
	where $R$ represents the Ricci scalar, $\kappa =8\pi G/c^3$ is a constant related to the three-dimensional Newtonian constant $G$, $\xi$ denotes the coupling constant with the dimension of $M^{-1}$, and $B_{\mu}$ is the bumblebee vector field with the strength $B_{\mu\nu}=\partial_{\mu}B_{\nu}-\partial_{\nu}B_{\mu}$. The potential $V$ of the bumblebee field has a minimum at $B_{\mu}B^{\mu}\pm b^2=0$, yielding a nonzero vacuum expectation value $\langle B_{\mu}\rangle=b_{\mu}$ with $b_{\mu}b^{\mu}=\mp b^2$ and leading to the violation of the $U(1)$ symmetry, where $b$ is a real positive constant and the signs ``$\pm$" mean that $b_{\mu}$ is timelike or spacelike. 
	
	By varying the action (\ref{action}) with respect to the metric, the authors of Ref. \cite{Ding:2023niy} obtained a rotating BTZ-like black hole solution 
	\begin{equation}\label{spacetime}
		ds^2 = -f(r) dt^2 +\frac{1+s}{f(r)} dr^2 + r^2 \left(d\varphi - \frac{j} {2r^2}dt\right)^2,
	\end{equation}
	with
	\begin{equation}
		f(r)=\frac{r^2}{\ell^2} - M +\frac{j^2}{4r^2},
	\end{equation}
	where $M$ and $j$ are the mass and spin parameters of the black hole, $s$ is the spontaneous Lorentz symmetry breaking parameter, and $\ell$ is the AdS radius. The determinant of the metric and the Kretschmann scalar may be obtained as
	\begin{equation}
		g=-(s+1)r^2\;,\;\;\;R^{\mu \nu \rho \lambda} R_{\mu \nu \rho \lambda}=\frac{12}{\ell^4 (1+s)^2},
	\end{equation}
	which means that the spacetime (\ref{spacetime}) has a singularity when $s=-1$. To keep the spacetime which is Lorentzian and is free of curvature singularity we, therefore, require the constraint $s>-1$. The black hole solution (\ref{spacetime}) has an event horizon and an inner horizon, $i.e.$, 
	\begin{equation}
		r^{2}_{\pm} = \frac{\ell^2}{2}\left(M \pm \sqrt{M^2 -\frac{j^2}{\ell^2}}\right),
	\end{equation} 
	which are exactly the same with the BTZ black hole solution and, therefore, are independent of the spontaneous Lorentz symmetry breaking parameter $s$. For the extremal BTZ-like black hole, we have $M^2 =j^2/\ell^2$ which leads to the coincidence of the event and inner horizons, $i.e.$, $r_+ =r_- =\ell \sqrt{M/2}$. As shown in Appendix A, we observe that extremal BTZ-like black holes do \textit{not} support stationary scalar clouds. Thus, we will focus on the non-extremal BTZ-like black holes in the following.
	
	In addition, the rotating BTZ-like black hole has an ergoregion $r_{erg} =\ell\sqrt{M}$ and the angular velocity of the horizon
	\begin{equation}\label{angularmom}
		\Omega_{H} =\frac{r_-}{\ell r_+},
	\end{equation}
	and the black hole mass parameter may be rewritten as 
	\begin{equation}\label{exeq}
		M = \frac{r^2_+ + r^2_-}{\ell^2} = \frac{r^2_+}{\ell^2} \left( 1 + \ell^2 \Omega^2_{H} \right),
	\end{equation}
	which are, again, exactly the same with the BTZ case.
	%%%
	%%%	
	\subsection{Massive Klein-Gordon equation}
	
	In the rotating BTZ-like black hole background, a massive scalar field $\Phi$, with the mass $\mu/l$, evolves according to the Klein-Gordon equation
	\begin{equation}\label{Klein-GordonEQ}
		\frac{1}{\sqrt{-g}}\frac{\partial}{\partial x^{\mu}}\left( \sqrt{-g} g^{\mu\nu} 
		\frac{\partial}{\partial x^{\nu}}\right) \Phi -\frac{\mu ^2}{\ell^2} \Phi  = 0,
	\end{equation}
	where $\mu$ is dimensionless. The spacetime isometries allow full seperation of variables, and by taking the ansatz
	
	\begin{equation}\label{ansatz}
		\Phi(t,r,\varphi) = e^{-i\omega t + ik\varphi} \phi(r),
	\end{equation}
	the radial equation of massive scalar fields turns into  
	\begin{equation}\label{kgeq}
		\frac{d^2 \phi(r)}{dr^2} +\left[\frac{1}{r} +\frac{1}{f(r)}\frac{df(r)}{dr}\right]\frac{d\phi(r)}{dr} +\frac{1+s}{f(r)} \left[-\frac{k^2}{r^2} -\frac{\mu^2}{\ell^2} +\frac{(j k-2r^2 \omega )^2}{4r^4 f(r)}\right] \phi(r) =0,
	\end{equation}
	where $\omega$ and $k$ are the frequency and the angular quantum number of scalar fields.	
	
	To solve the radial equation~\eqref{kgeq}, one may introduce a new radial coordinate $z$ for non-extremal BTZ-like black holes as
	\begin{equation}\label{z1}
		z\equiv \frac{r^2-r^2_+}{r^2-r^2_-},\qquad r_+ \ne r_-,
	\end{equation}
	which transforms the exterior region from $r\in (r_+,+\infty)$ to $z \in (0,1)$, and Eq. (\ref{kgeq}) becomes
	\begin{equation}\label{noneq}
		z(1-z)\frac{d^2 \phi(z)}{dz^2} +(1-z)\frac{d \phi(z)}{dz} +\left(\frac{A}{z}+B+\frac{C}{1-z}\right)\phi(z)=0  ,
	\end{equation}
	with
	\begin{equation}
		A= \frac{\ell^4(\omega r_+ -kr_{-}/\ell)^2}{4(r^2_+-r^2_-)^2}(1+s) ,\;\; B =-\frac{\ell^4(\omega r_- -kr_{+}/\ell)^2}{4(r_+^2 -r_-^2)^2}(1+s), \;\; C=-\frac{\mu^2(1+s)}{4}.
	\end{equation}
	Then, letting $\phi(z) = z^{\alpha}(1-z)^{\beta} F(z)$, one obtains the standard hypergeometric equation~\cite{AbramowitzStegun}
	\begin{equation}
		z(1-z)\frac{d^2F(z)}{dz^2} +\left[c -(1+a+b)z\right]\frac{dF(z)}{dz} -ab F(z)=0,
	\end{equation}
	with
	\begin{equation}\label{abc}
		\begin{split}
			a\equiv  &\beta -i\ell \frac{(\omega \ell + k)\sqrt{1+s}}{2(r_+ +r_-)},\qquad b\equiv \beta - i\ell \frac{(\omega \ell- k)\sqrt{1+s}}{2(r_+ - r_-)},\qquad c\equiv 1+2\alpha,\\
			\alpha \equiv &-i \frac{\ell^2r_+}{2(r^2_+-r^2_-)}(\omega -k \Omega_{H})\sqrt{1+s},\qquad \beta \equiv \frac{1}{2}\left[1+\sqrt{1+\mu^2(1+s)}\right],
		\end{split}
	\end{equation}
	where $F(z)$ is the Gaussian hypergeometric function.
	
	\subsection{Boundary conditions}
	
	To look for scalar clouds and scalar quasinormal modes by solving Klein-Gordon equation, we have to employ an ingoing wave boundary condition at the event horizon and the Robin type vanishing energy flux boundary condition at infinity, following the Ref. \cite{WangAU1,WangAU2,WangPRD2015,Ferreira:2017cta,Dappiaggi:2017pbe}.
	
	At the horizon $z=0$, one obtains the general solution for $\phi(z)$
	\begin{equation}
		\phi(z)= A z^{\alpha} (1-z)^{\beta} F(a,b;c;z)+B z^{-\alpha}(1-z)^{\beta}F(a-c+1,b-c+1;2-c;z),
	\end{equation}
	where $A$ and $B$ are two integration constants, and an ingoing wave boundary condition implies $B=0$.
	
	At infinity $z \to 1$, the general solution of $\phi(z)$ is given by
	\begin{equation}\label{solution}
		\begin{split}
			\phi(z) =&C^{(D)}\phi^{(D)}(z)+C^{(N)}\phi^{(N)}(z)\\
			=&A\left[\frac{\Gamma(c)\Gamma(c-a-b)}{\Gamma(c-a)\Gamma(c-b)}\phi^{(D)}(z)+
			\frac{\Gamma(c)\Gamma(a+b-c)}{\Gamma(a)\Gamma(b)}\phi^{(N)}(z)\right],
		\end{split}
	\end{equation}
	with two linearly independent solutions
	\begin{equation}\label{solution2}
		\begin{split}
			\phi^{(D)}(z) =&z^{\alpha} (1-z)^{\beta} F(a,b;a+b-c+1;1-z),\\ 
			\phi^{(N)}(z) =&z^{\alpha} (1-z)^{1-\beta} F(c-a,c-b;c-a-b+1;1-z),
		\end{split}
	\end{equation}
	where $C^{(D)}$ and $C^{(N)}$ are two complex constants, and the parameters $a$, $b$, $c$, $\alpha$ and $\beta$ are given in Eq.~\eqref{abc}. Here $\phi^{(D)}(z)$ and $\phi^{(N)}(z)$ are introduced to denote the Dirichlet and Neumann solutions \cite{Ferreira:2017cta}, and to keep these two solutions square-integrable near infinity, we will concentrate on the parameter space
	\begin{equation}
		-1< s <-\left(1+\frac{1}{\mu^2}\right)\;,\;\;\;\mu^2 <0.
	\end{equation}
	
	By calculating the flux of energy at $r \to \infty$ ($z \to 1$)
	\begin{equation}
		\mathcal{F}=\lim\limits_{r\to\infty}\int_{\sum_r}d\phi \sqrt{-g}g^{rr}T_{rt}\;,	
		\end{equation}
	where $\sum_r$ and $T_{\mu\nu}$ are the hypersurface of constant $r$ and the energy momentum tensor of scalar fields, the vanishing of energy flux \cite{WangPRD2021,WangEPJC2021,Ferreira:2017cta} leads to
	\begin{equation}\label{Phisincos}
		\phi=\cos(\zeta)\phi^{(D)}+\sin(\zeta)\phi^{(N)},
	\end{equation}
	where the boundary parameter $\zeta$ satisfies $\zeta\in\left[0,\pi\right)$. It is shown clearly that $\zeta=0$ corresponds to the Dirichlet boundary condition and $\zeta=\pi/2$ corresponds to the Neumann boundary condition. 
	
	By comparing Eq. (\ref{solution}) with Eq.  (\ref{Phisincos}), we have the boundary condition
	\begin{equation}\label{bc}
		\tan\zeta=\dfrac{\Gamma(a+b-c)\Gamma(c-a)\Gamma(c-b)}{\Gamma(c-a-b)\Gamma(a)\Gamma(b)}\;.
	\end{equation}
	
	\section{Stationary scalar clouds}
	
	In this section, we investigate stationary scalar configurations around the rotating BTZ-like black hole in the Einstein-bumblebee gravity. For this case, the resonance condition should be satisfied, $i.e.$, the field's frequency equals to the critical frequency
	\begin{equation}\label{resonance}
		\omega=\omega_c = k \Omega_{H},
	\end{equation}
	which leads to $\alpha = 0$ and $c = 1$, $a = \beta - i \ell \sqrt{1+s} k/(2r_+)$ in Eq. (\ref{abc}). Then the solutions in Eq. (\ref{solution2}) reduce to 
	\begin{eqnarray}
		\phi^{(D)}(z)&= & (1-z)^{\beta} F(a,a^*;2\beta ;1-z),\\ \nonumber
		\phi^{(N)}(z)&= & (1-z)^{1-\beta} F(1-a,1-a^*;2-2\beta ;1-z),
	\end{eqnarray}
	where $\beta$ is given in Eq. (\ref{abc}), and the boundary condition~\eqref{bc} for scalar clouds becomes
	\begin{equation}\label{ResonanceCondition}
		\begin{split}
			\tan(\zeta)=\frac{\Gamma(2\beta-1)\Gamma(1-a)\Gamma(1-a^*)}{\Gamma(1-2\beta)\Gamma(a)\Gamma(a^*)} 
			=\frac{\Gamma(\sqrt{1+(1+s)\mu^2})|\Gamma(\frac12-\frac12\sqrt{1+(1+s)\mu^2}+i\frac{k\ell \sqrt{1+s}}{2r_+})|^2}
			{\Gamma(-\sqrt{1+(1+s)\mu^2})|\Gamma(\frac12+\frac12\sqrt{1+(1+s)\mu^2}+i\frac{k\ell \sqrt{1+s}}{2r_+})|^2}.
		\end{split}
	\end{equation}
	
	By scanning the parameter space of the system for given values of $s$, $\mu^2$, $k$, $r_+$ and $\zeta$, one may obtain scalar clouds. Here, for concreteness we take $\mu^2=-0.65$, $\zeta=0.9\pi$ and, without loss of generality, we take the AdS radius $\ell=1$ to measure all other physical quantities. One should further note that the boundary parameter $\zeta$ is bounded in the range $[\zeta_*,\pi)$, where
	\begin{eqnarray}
		\zeta_*=\arctan \left(\frac{\Gamma(\sqrt{1+(1+s)\mu^2})|\Gamma(\frac12-\frac12\sqrt{1+(1+s)\mu^2})|^2}
		{\Gamma(-\sqrt{1+(1+s)\mu^2})|\Gamma(\frac12+\frac12\sqrt{1+(1+s)\mu^2})|^2}\right),
	\end{eqnarray}
	satisfying $\zeta_* \in(\frac{\pi}{2},\pi)$. Since $\zeta_*$ increases as the Lorentz symmetry breaking parameter $s$ increases, the allowed parameter space decreases as the increase of $s$.
	
\begin{figure}[htbp]
		\subfigure{\includegraphics[scale=0.92]{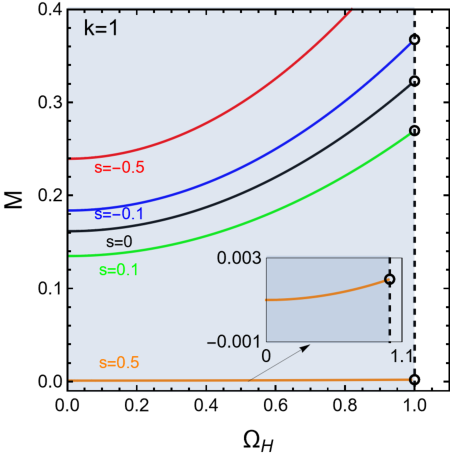}}\hspace{8ex}
		\subfigure{\includegraphics[scale=0.92]{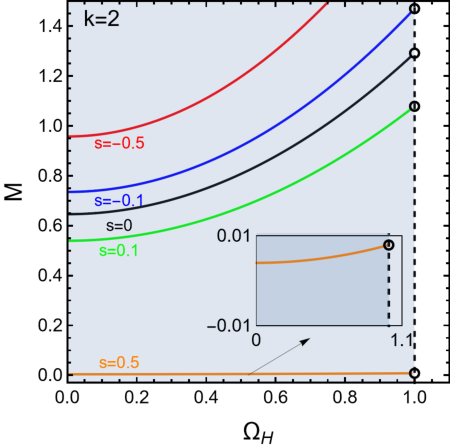}}\\ \vspace*{4ex}
		\caption{Existence lines of nodeless scalar clouds ($n=0$) for various Lorentz symmetry breaking parameter $s$ with fixed quantum numbers $k=1$ (left), $k=2$ (right), in the mass vs horizon angular velocity parameter space of BTZ-like black holes. The dashed black curve corresponds to extremal BTZ-like black holes with $\Omega_H =1/\ell$ and non-extremal BTZ-like black holes exist in the shaded region.}\label{ks1}
	\end{figure}

\begin{figure}[htbp]
		\subfigure{\includegraphics[scale=0.90]{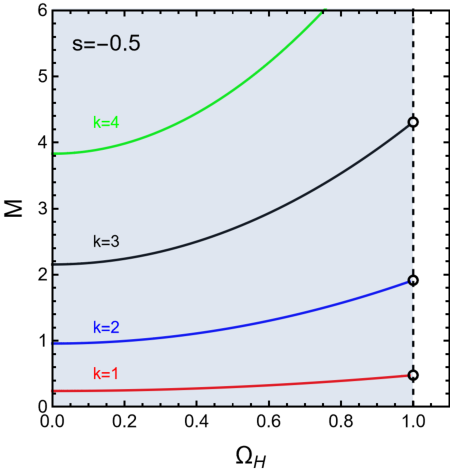}}\hspace{8ex} 
		\subfigure{\includegraphics[scale=0.97]{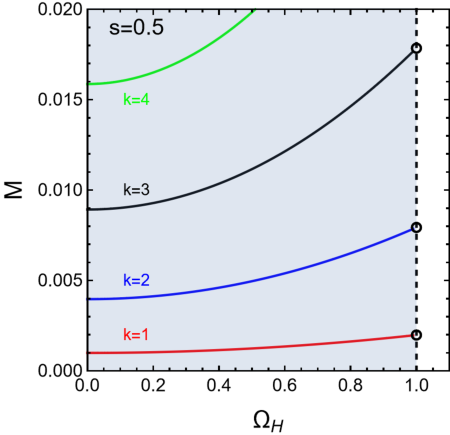}}\vspace*{4ex}
		\caption{Existence lines of nodeless scalar clouds ($n=0$) for various quantum number $k$ with fixed $s=-0.5$ (left), $s=0.5$ (right), in the mass vs horizon angular velocity parameter space of BTZ-like black holes. The dashed black curve corresponds to extremal BTZ-like black holes with $\Omega_H =1/\ell$ and non-extremal BTZ-like black holes exist in the shaded region.}\label{ks2}
	\end{figure}

	To explore the impact of the Lorentz symmetry breaking parameter $s$ on scalar clouds, in Fig.~\ref{ks1} we present scalar clouds for the case of $n=0$, $\zeta= 0.9\pi$ and for various Lorentz symmetry breaking parameters $s$ ($s=-0.5, -0.1, 0, 0.1, 0.5$) with fixed $k=1$ (left panel) and $k=2$ (right panel). It is shown that the clouds exist for smaller background mass with fixed background angular velocity by increasing $s$ with fixed $k$. In Fig.~\ref{ks2} we exhibit scalar clouds for the case of $n=0$, $\zeta= 0.9\pi$ and for various quantum number $k$ ($k=1, 2, 3, 4$) with fixed $s=-0.5$ (left panel) and $s=0.5$ (right panel). It is shown that the clouds exist for larger background mass with fixed background angular velocity by increasing $k$ with fixed $s$. The results shown in the above figures indicate the Lorentz symmetry breaking parameter $s$ and the quantum number $k$ have entirely opposite effects on the existence lines of nodeless scalar clouds. One should further notice that the Lorentz symmetry breaking parameter $s$ has no effect on the extremal lines (the dashed black curve in the figure), which is consistent with Eq.~\eqref{angularmom}. This is completely different from the effect of the MOG parameter $\alpha$ on stationary scalar clouds in Kerr-MOG black holes, where the increase of parameter $\alpha$ makes the extremal lines lower \cite{Qiao:2020fta}. 
	
	%%%%%%
	%%%%%%
\begin{figure}[htbp]
		\subfigure{\includegraphics[scale=1]{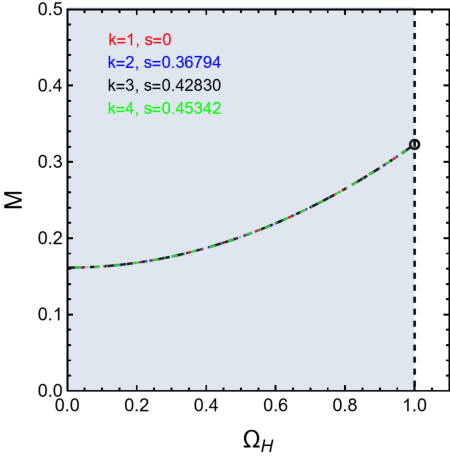}}
		\caption{Degenerate existence lines of nodeless scalar clouds ($n=0$) for several pairs of the quantum number $k$ and Lorentz symmetry breaking parameters $s$, in the mass vs horizon angular velocity parameter space of BTZ-like black holes, with $\mu^2=-0.65$, $r_+=0.40182$, $\ell=1$ and with the Robin boundary condition $\zeta= 0.9\pi$ at infinity.}\label{dege}
	\end{figure}

\begin{figure}[htbp]
		\subfigure{\includegraphics[scale=1]{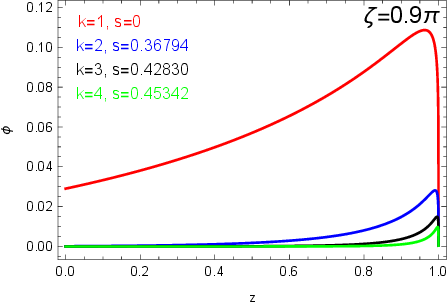}} \vspace*{4ex}
		\caption{The radial profile of degenerate clouds on BTZ-like black holes is presented in terms of $z$, with $\mu^2=-0.65$, $r_+=0.40182$, $\ell=1$ and with the Robin boundary condition $\zeta= 0.9\pi$ at infinity.}\label{Phiz}
	\end{figure}

	%%%
	Based on the above observation, in Fig. \ref{dege} we present scalar clouds in the parameter space by taking a few pairs of ($k, s$), including $(k=1, s=0)$, $(k=2, s=0.36794)$,  $(k=3, s=0.42830)$ and $(k=4, s=0.45342)$. It is interesting to note that the existence lines for all the cases we considered here are the same, and we term such configurations as degenerate clouds. This indicates that infinite degenerate clouds may be constructed by increasing $k$ and the corresponding $s$ is bounded by $s=-(1+1/\mu^2)$. The existence of degenerate clouds raises a question, $i.e.$ how to distinguish BTZ black holes and BTZ-like black holes. To solve this issue, we further explore the radial profile of degenerate clouds in Fig. \ref{Phiz}, by taking the exactly the same parameters given in Fig. \ref{dege}. It is shown clearly that the radial profiles of degenerate clouds $\phi$ are completely different, indicating that degenerate clouds may be discerned by the radial profile.
	%%%
	%%%	
	%%%
	
	Before proceeding, we would like to stress that in the above we only find the nodeless scalar clouds ($n=0$), $i.e.$, the ground state stationary clouds, which is different from the Kerr case \cite{Herdeiro:2014goa,Wang:2018xhw}. Considering that the clouds only exist at the threshold of superradiance \cite{Brito:2015oca}, we will study the quasinormal modes in the BTZ-like black hole and find out the corresponding reasons.
	
	\section{Quasinormal modes}\label{IVC}
	
	For a massive scalar field $\Phi$ satisfying the Klein-Gordon equation (\ref{Klein-GordonEQ}), from Eqs. (\ref{solution}) and  (\ref{Phisincos}) we can obtain quasinormal frequencies by solving the following equation
	\begin{equation}\label{Quasi-boundState}
		\begin{split}
			\tan(\zeta)
			=&\frac{\Gamma\left(\sqrt{1+(1+s)\mu^2}\right)\Gamma\left(\frac12-\frac12\sqrt{1+(1+s)\mu^2}-i\ell \sqrt{1+s}\frac{\omega\ell-k}{2(r_+ -r_-)}\right)}{\Gamma\left(-\sqrt{1+(1+s)\mu^2}\right)\Gamma\left(\frac12+\frac12\sqrt{1+(1+s)\mu^2}-i\ell \sqrt{1+s}\frac{\omega\ell+k}{2(r_+ +r_-)}\right)}\\
			&\times\frac{\Gamma\left(\frac12-\frac12\sqrt{1+(1+s)\mu^2}-i\ell \sqrt{1+s}\frac{\omega\ell+k}{2(r_+ + r_-)}\right)}{\Gamma\left((\frac12+\frac12\sqrt{1+(1+s)\mu^2}-i\ell \sqrt{1+s}\frac{\omega\ell-k}{2(r_+-r_-)}\right)},
		\end{split}
	\end{equation}
	where the Robin parameter satisfies $\zeta\in[0,\pi)$. Obviously, if $\zeta=0$, $i.e.$, the so-called Dirichlet boundary condition,  we can solve Eq. (\ref{Quasi-boundState}) exactly, and get the left-moving frequencies
	\begin{equation}
		\omega ^{(D),L} =-\frac{k}{\ell} - i \frac{r_+ + r_-}{\sqrt{1+s} ~\ell^2} \left[2n+1+ \sqrt{1+ \mu ^2 (1+s)}\right] ,
	\end{equation} \label{omegaDL}
	and the right-moving frequencies
	\begin{equation} 
		\omega ^{(D),R} = \frac{k}{\ell} - i \frac{r_+ - r_-}{\sqrt{1+s} ~\ell^2} \left[2n+1+ \sqrt{1+ \mu ^2 (1+s)}\right],
	\end{equation}\label{omegaDR}
	which reduce to the quasinormal frequencies given in Ref. \cite{Dappiaggi:2017pbe} for the BTZ black hole if the Lorentz symmetry breaking parameter $s=0$. Similarly, if $\zeta=\pi/2$, $i.e.$, the so-called Neumann boundary condition, we have
	\begin{eqnarray}\label{omegaN}
		\omega ^{(N),L} &=&-\frac{k}{\ell} - i \frac{r_+ + r_-}{\sqrt{1+s} ~\ell^2} \left[2n+1 - \sqrt{1+ \mu ^2 (1+s)}\right], \\
		\omega ^{(N),R} &=& \frac{k}{\ell} - i \frac{r_+ - r_-}{\sqrt{1+s} ~\ell^2} \left[2n+1 - \sqrt{1+ \mu ^2 (1+s)}\right].
	\end{eqnarray}
	For other values of Robin parameter $\zeta$, we have to solve Eq. (\ref{Quasi-boundState}) numerically and obtain the numerical results for the quasinormal frequencies.	
	\begin{figure}[htbp]
		\subfigure{\includegraphics[scale=0.95]{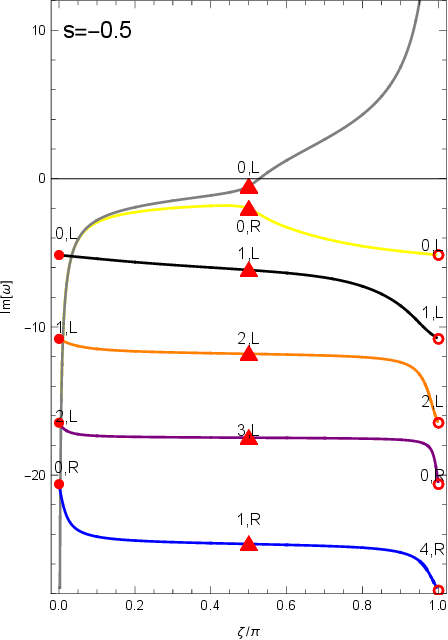}}\hspace{8ex}
		\subfigure{\includegraphics[scale=0.95]{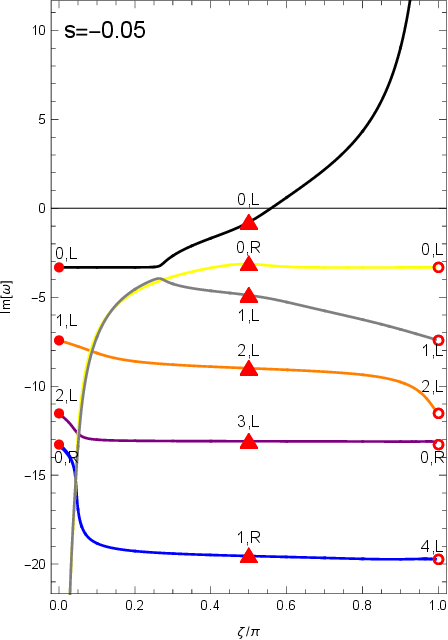}}\\ \vspace*{4ex}
		\subfigure{\includegraphics[scale=0.95]{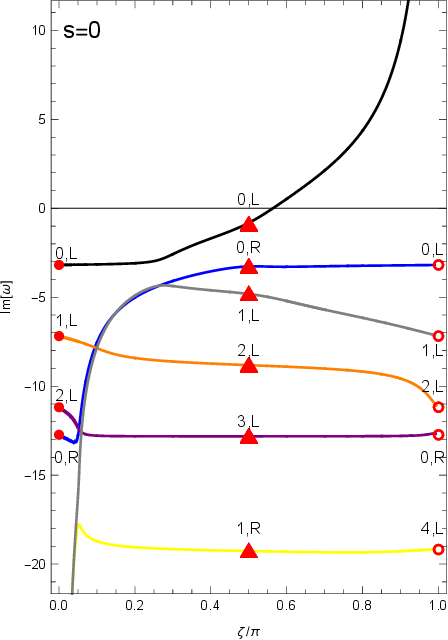}} \hspace{8ex}
		\subfigure{\includegraphics[scale=0.95]{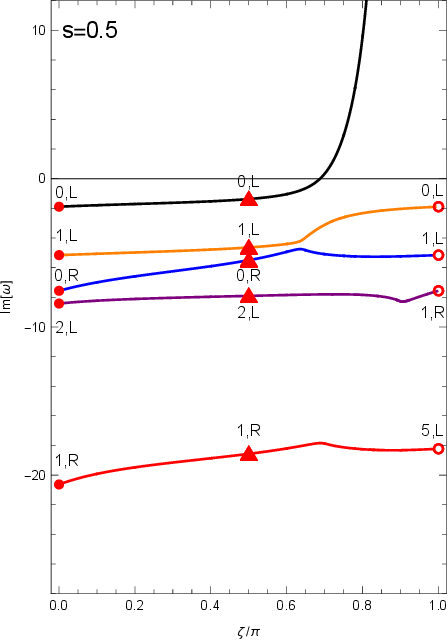}}
		\caption{Imaginary parts of some quasinormal frequencies as a function of $\zeta/ \pi$ with different Lorentz symmetry breaking parameters $s$ for BTZ-like black holes and scalar field with $\mu^2=-0.65$, $r_+=5$, $r_-=3$, $\ell=1$ and $k=1$.}\label{qusi1}
	\end{figure} 
	
	In Fig. \ref{qusi1}, we present the imaginary parts of some selected quasinormal frequencies as a function of $\zeta/ \pi$ with different Lorentz symmetry breaking parameters for a given black hole configuration (the fixed $\ell$, $r_+$ and $r_-$) and scalar field configuration (the fixed $\mu^2$ and $k$). The red dots denote quasinormal frequencies corresponding to the Dirichlet boundary condition $(\zeta=0)$ and the red triangles correspond to those for the Neumann boundary condition $(\zeta=\frac{\pi}{2})$, which can be served as a check of the numerical computation. We can see clearly that, as a function of $\zeta$, the mode continuously connects the case of Dirichlet boundary condition with that of Neumann boundary condition for quasinormal frequencies. However, all the imaginary parts of quasinormal frequencies are negative except for the fundamental frequencies of type L, corresponding to the nodeless modes ($n=0$). For the nodeless modes, the imaginary parts of frequencies always change from negative to positive as the parameter $\zeta$ increases, which means that there exist the so-called superradiance zero-modes, occurring in between decaying modes and superradiantly amplified modes \cite{Ferreira:2017cta}. It should be noted that, regardless of the Lorentz symmetry breaking parameter $s$, the superradiance zero-modes only appear in the nodeless modes ($n=0$) of type L, which indicates that the stationary scalar clouds only exist in this ground state since the resonance condition corresponds precisely to the threshold of the superradiant instability \cite{Brito:2015oca}. This supports the observation shown in Figs. \ref{ks1}, \ref{ks2} and \ref{dege} where we can only obtain the existence lines of nodeless scalar clouds ($n=0$).	
	
	\begin{figure}[htbp]
		\includegraphics[scale=0.95]{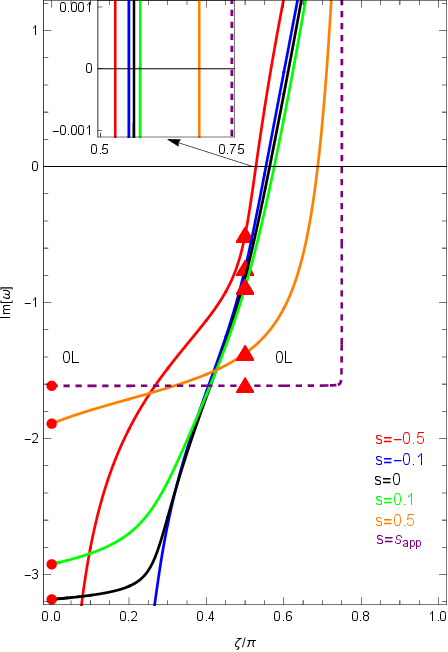}
		\caption{Imaginary parts of some quasinormal frequencies of nodeless modes ($n=0$) as a function of $\zeta/ \pi$ with different Lorentz symmetry breaking parameters $s$ for BTZ-like black holes and scalar field with $\mu^2=-0.65$, $r_+=5$, $r_-=3$, $\ell=1$ and $k=1$. The dashed purple curve corresponds to the case of $s_{app}=0.53846$. 
		}\label{qusi2}
	\end{figure}
	
	In order to further analyze the superradiance zero-modes, we display the imaginary parts of some quasinormal frequencies of nodeless modes ($n=0$) as a function of $\zeta/ \pi$ with different Lorentz symmetry breaking parameters for a given black hole configuration (the fixed $\ell$, $r_+$ and $r_-$) and scalar field configuration (the fixed $\mu^2$ and $k$) in Fig. \ref{qusi2}. We observe that the threshold of the superradiant instability, $i.e.$, the critical value of $\zeta$ with Im$[\omega]=0$, increases as the Lorentz symmetry breaking parameter $s$ increases. In particular, we find this threshold of the superradiant instability is identically equal to the value of $\zeta$ given by Eq. (\ref{ResonanceCondition}) for the existence of the stationary scalar clouds. From Fig. \ref{qusi2}, we note that the critical value of $\zeta\rightarrow 0.75\pi$ when $s=s_{app}$. Thus, we argue that, in the case of $\mu^2=-0.65$, $r_+=5$, $r_-=3$, $\ell=1$ and $k=1$, the threshold of the superradiant instability has a range $0.5\pi < \zeta < 0.75\pi$, which suggests that the stationary scalar clouds only exist in this range. As a matter of fact, we only take $\mu^2=-0.65$ and $\ell=1$ in Figs.~\ref{ks1}-\ref{Phiz} with different valves of $r_+$, and so the superradiant instability can appear in the case of $\zeta=0.9\pi$.

	\section{Conclusions}\label{Sec.V}
	
	We have investigated the Klein-Gordon equation for a massive scalar field on a rotating BTZ-like black hole in the Einstein-bumblebee gravity. By employing appropriate Robin boundary conditions at the AdS boundary, we have established the existence of stationary fundamental clouds. We found that, by increasing the Lorentz symmetry breaking parameter $s$ (the quantum number $k$), the clouds exist for smaller (larger) background mass with the same background angular velocity, which means that the parameters $s$ and $k$ have entirely opposite effects on the existence lines of scalar clouds. In particular, we obtained the degenerate clouds where different parameters $s$ and $k$ have the same existence line of scalar clouds, which has not ever been observed as far as we know. Interestingly, we pointed out that the radial profiles of degenerate clouds $\phi$ may be used to discriminate degenerate clouds. Moreover, we noticed that, below the limiting value $s=-(1+1/\mu^2)$, there exist infinite degenerate clouds for any initial values of $s$ and $k$.
	
	By analyzing quasinormal modes in BTZ-like black holes, we observed that, regardless of the Lorentz symmetry breaking parameter $s$, superradiant instabilities are only appeared in the nodeless mode ($n=0$) of type L, which illustrates the observation, shown in Figs. \ref{ks1}, \ref{ks2} and \ref{dege}, that only fundamental clouds are allowed. Moreover, we observed that the threshold of the superradiant instability, which determines scalar clouds, depends on the parameter $s$. Our results indicate that the Lorentz symmetry breaking parameter $s$ yields richer physics in the stationary clouds of massive scalar fields around black holes.

	\begin{acknowledgments}		
		
This work was supported by the National Key Research and Development Program of China (Grant No. 2020YFC2201400), National Natural Science Foundation of China (Grant Nos. 12275079, 12475050 and 12035005), the Scientific Research Fund of Hunan Provincial Education Department (Grant No. 22A0039) and innovative research group of Hunan Province (2024JJ1006).
		
	\end{acknowledgments}
	
	\appendix
	
	\section{Massive Klein-Gordon equation in the extremal BTZ-like black hole}
	
	In this appendix, we consider the massive Klein-Gordon equation in the extremal BTZ-like black hole of the Einstein-bumblebee gravity. For this case, we have $r_+ =r_- =\ell \sqrt{M/2}$ since $M^2 =j^2/\ell^2$. Thus, the angular velocity of the event horizon is completely determined by the AdS radius, $\Omega_H =1/\ell$, and the black hole mass can be expressed as $M= 2 r_+^2 \Omega_H^2$. We still take the ansatz (\ref{ansatz}) but change the radial coordinate (\ref{z1}) into 
	\begin{equation}\label{z2}
		z \equiv \frac{r^2_+}{r^2 - r^2_+},
	\end{equation}
	which maps the exterior region from $r \in (r_+, +\infty)$ to $z\in (0, +\infty)$ with the AdS boundary at $z\to 0$. Then Eq. (\ref{kgeq}) turns into
	\begin{equation}
		\frac{d^2 \phi(z)}{dz^2} +\left(\frac{A}{z} +B +\frac{C}{z^2} \right) \phi(z) =0,
	\end{equation}
	with
	\begin{eqnarray}
		A=\frac{\ell^2(\omega^2\ell^2-k^2 )}{4r_+^2}(1+s),\;\;B=\frac{\ell^2(\omega\ell -k)^2}{4r_+^2}(1+s),\;\;C=-\frac{\mu^2(1+s)}{4}.
	\end{eqnarray}
	Thererfore, with the Kummer's confluent hypergeometric function $M(a,b,z)$, we get two linearly independent solutions
	\begin{equation}
		\begin{split}
			\phi^{(D)}(z) = &z^{\beta} e^{i \alpha z} M(a,b,-2i\alpha z),\\ 
			\phi^{(N)}(z) = &z^{1-\beta}e^{i \alpha z} M(a-b+1,2-b,-2i\alpha z),
		\end{split}
	\end{equation}
	where $\beta$ is still as in Eq. (\ref{abc}) and
	\begin{equation}
		\alpha \equiv \ell \frac{\omega\ell - k}{2r_+}\sqrt{1+s},\qquad  a\equiv \beta -i\ell \sqrt{1+s}\frac{\omega\ell+k}{4r_+},\qquad b=2\beta.
	\end{equation}
	
	In order to obtain stationary scalar modes, we impose the resonance condition $\omega=k \Omega_{H} =k/\ell$ and arrive at
	\begin{equation}
		\phi^{(D)}(z) = z^{\beta},\qquad \phi^{(N)}(z) = z^{1-\beta}.
	\end{equation}
	Obviously, in the extremal case, there is no linear combination of the solutions which is regular at the horizon $z\to +\infty$. Thus, just as in the extremal BTZ case \cite{Ferreira:2017cta}, one can not find stationary scalar cloud configurations around the extremal BTZ-like black hole.

	% BIBLIOGRAPHY

\end{document}